\documentclass[sigconf, authorversion]{acmart}

\AtBeginDocument{%
  }

\copyrightyear{2025}
\acmYear{2025}
\setcopyright{rightsretained}
\acmConference[RecSys '25]{Proceedings of the Nineteenth ACM Conference on Recommender Systems}{September 22--26, 2025}{Prague, Czech Republic}
\acmBooktitle{Proceedings of the Nineteenth ACM Conference on Recommender Systems (RecSys '25), September 22--26, 2025, Prague, Czech Republic}\acmDOI{10.1145/3705328.3748136}
\acmISBN{979-8-4007-1364-4/2025/09}

\begin{document}

\title{Location Matters: Leveraging Multi-Resolution Geo-Embeddings for Housing Search}


\author{Ivo Silva}
\email{ivo.silva@quintoandar.com.br}
\affiliation{%
  \institution{QuintoAndar}
  \city{Lisbon}
  \country{Portugal}
}

\author{Pedro Nogueira}
\email{pedro.nogueira@quintoandar.com.br}
\affiliation{%
  \institution{QuintoAndar}
  \city{Lisbon}
  \country{Portugal}
}

\author{Guilherme Bonaldo}
\email{guilherme.bonaldo@quintoandar.com.br}
\affiliation{%
  \institution{QuintoAndar}
  \city{Lisbon}
  \country{Portugal}
}

\renewcommand{\shortauthors}{Silva et al.}

\begin{abstract}
QuintoAndar Group is Latin America’s largest housing platform, revolutionizing property rentals and sales. Headquartered in Brazil, it simplifies the housing process by eliminating paperwork and enhancing accessibility for tenants, buyers, and landlords. With thousands of houses available for each city, users struggle to find the ideal home. In this context, location plays a pivotal role, as it significantly influences property value, access to amenities, and life quality. A great location can make even a modest home highly desirable. Therefore, incorporating location into recommendations is essential for their effectiveness. We propose a geo-aware embedding framework to address sparsity and spatial nuances in housing recommendations on digital rental platforms. Our approach integrates an hierarchical H3 \cite{uber_h3} grid at multiple levels into a two-tower neural architecture. We compare our method with a traditional matrix factorization baseline and a single-resolution variant using interaction data from our platform. Embedding specific evaluation reveals richer and more balanced embedding representations, while offline ranking simulations demonstrate a substantial uplift in recommendation quality.
\end{abstract}




\maketitle

\section{Introduction}
The emergence of digital housing platforms has transformed how renters search for and discover available properties. These platforms enable advanced recommender systems to surface the most relevant listings. Accurate house location recommendations are essential for the effectiveness of these systems. To achieve this, two key challenges must be addressed. First, the number of houses available in any given location is inherently limited and unevenly distributed, leading to severe item sparsity when making recommendations within narrow geographic boundaries. Second, user search behavior often extends beyond strictly adjacent neighborhoods, as renters frequently consider locations that share key attributes, such as proximity to the beach, healthcare facilities, public transit, safety levels, and price points.
To quantify the importance of location, we conducted a survey with nearly 800 respondents, finding that more than 50\% of the participants rank location as the most critical factor in their rental decision-making process. Although there is a clear user preference for location, our recommender suboptimally recognizes the importance of this concept, even with access to several geographic features. Due to this, we fail to capture the rich spatial context that influences user choice, often resulting in recommendations for properties in locations that do not align with user intent.
The main contributions of this work are as follows:
\begin{enumerate}
    \item We propose a method for generating robust embeddings that capture location information at multiple levels of granularity.
    \item We comprehensively evaluate these embeddings through two complementary approaches: (a) intrinsic metrics focused on embedding quality, and (b) extrinsic evaluation using real-world production logs to measure practical impact on recommendation performance.
\end{enumerate}

\section{Methodology}
In this section, we detail our systematic approach to the construction and evaluation of geo-aware embeddings. We first compare the traditional matrix factorization \cite{matrix_factorization_paper}  model to a two-tower neural network trained using contrastive loss \cite{two_tower_paper}. To enhance spatial awareness, we extend the two-tower model by integrating multiresolution H3 embeddings, effectively capturing spatial signals at varying granularities. We detail the feature sets, architectural designs, and training objectives guiding our experimentation, followed by specific implementation insights. Our goal is to develop robust embedding methods that accurately integrate user preferences with geographic context.

\subsection{Data and Features}
We train our models on user-house interactions collected through the platform from 1 January 2022 to 31 December 2024. We select only high-intent events, like booking a visit or making an offer. Lower intent signals, such as clicks, are excluded. Our feature set includes:
\begin{itemize}
    \item {\bfseries User ID}
    \item {\bfseries City}
    \item {\bfseries H3 cell indices at resolutions 6–9.} H3 (Hexagonal Hierarchical Spatial Index) converts latitude/longitude pairs into compact 64-bit hexagonal identifiers at multiple resolutions, ranging from coarse (\textasciitilde 43 km² at res 6) to fine (\textasciitilde 0.01 km² at res 9), providing a uniform, neighbor-consistent grid for multiscale geographic context.
\end{itemize}

\subsection{Modeling Approaches}
We compare three methods for learning geo-aware embeddings in our house-search recommender. We will detail each method in the subsequent sections.

\subsubsection{Matrix Factorization}
As a classical baseline we adopt the implicit-feedback matrix factorization model. The method represents each user and each location with a low-dimensional latent vector and learns these vectors with Alternating Least Squares (ALS). The only input feature for each item is its finest-grained H3 cell identifier (resolution 9), leaving all geographic information to emerge from the learned factors.

\subsubsection{Two-tower Model (H3 Resolution 9)}
The two-tower baseline embeds only the H3 resolution-9 cell for locations and the user ID for users, each via an embedding layer and a few dense layers. This minimal setup serves as a baseline for assessing richer geographic inputs.

\subsubsection{Two-tower Multi-Resolution Model}
\begin{figure}[h]
  \centering
  \includegraphics[width=\linewidth]{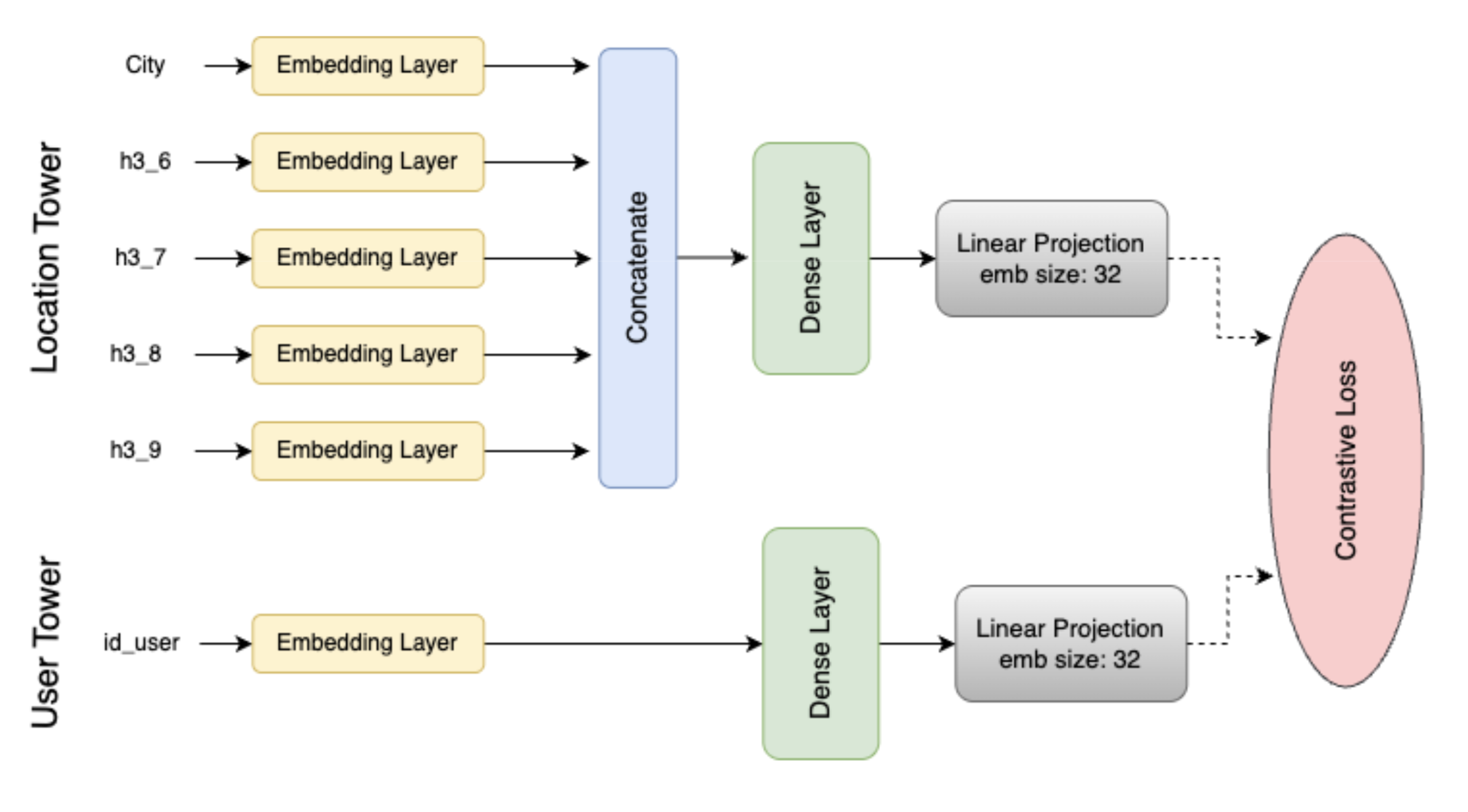}
  \caption{Two-tower multi-resolution model architecture.}
  \label{twotower}
\end{figure}

Figure~\ref{twotower} illustrates our enhanced two-tower architecture, which enriches geographic context in two steps:
\begin{itemize}
  \item \textbf{City Embedding.} We map each city name through a dedicated embedding layer to capture high-level urban features.
  \item \textbf{Multi-Resolution H3 Embeddings.} We embed H3 cells at resolutions 6, 7, 8 and 9, each via its own layer, to capture spatial signals from broad regions down to fine detail.
\end{itemize}
We then concatenate the city and all H3 embeddings into one vector and refine it through two fully connected layers with ReLU activations, yielding the final location embedding. The user tower follows the same design as in Section 2.2.2.

\subsection{Training Objective and Loss Function}
The two-tower models are trained to maximize the similarity between user and location embeddings for observed (positive) interactions, while minimizing similarity with unobserved (negative) pairs within the same batch. This is achieved using a contrastive loss based on the InfoNCE formulation \cite{oord2019representationlearningcontrastivepredictive}.

Additionally, a masking mechanism is applied during training to prevent the model from comparing user interactions with locations that are known positives but occur off-diagonal in the similarity matrix. This in-batch negative sampling is crucial to ensure that the embeddings are discriminative with respect to user interaction history.

\subsection{Implementation Details}
The model is implemented in PyTorch Lightning \cite{falcon_pytorch_lighting} to enable modular training and logging. We use Xavier uniform initialization for all embedding and linear weights, 32-dimensional embeddings, optimize with Adam (learning rate 0.001), set a batch size of 512 and train for 15 epochs.

\section{Evaluation}

In this section, we assess the quality of our geo-aware embeddings using both intrinsic and extrinsic evaluation protocols. We first present the intrinsic evaluation in Section 3.1, then the extrinsic evaluation in Section 3.2 and finally discuss the combined results in Section 3.3.

\subsection{Intrinsic Evaluation}

We create an embedding for every location (H3 cell) observed during training and run collapse-detection metrics on the complete set of vectors. To quantify embedding quality, we employ:

\begin{itemize}
  \item \textbf{Information Abundance}: as defined by Guo et al.\cite{guo2024embeddingcollapsescalingrecommendation}, which measures the diversity and richness of the encoded information.
  \item \textbf{Spectrum of Singular Values}: following Jianl et al.\cite{jing2022understandingdimensionalcollapsecontrastive}, we compute the singular values of the normalized embedding covariance matrix. A slower decay in the log‐spectrum indicates a more balanced distribution of variance across embedding dimensions.
\end{itemize}

\subsection{Extrinsic Evaluation}

We analyse one month of production logs from March 2025 from the listing similar houses module, an item-item recommender that shows each user a list of houses similar to an anchor house. For each anchor, we retrieve the original top-$N$ recommendations generated by the baseline and compute the cosine similarity between the geo-embedding of the anchor and each candidate’s embedding. We then conduct an ablation study using a set of similarity thresholds ($\tau$), progressively pruning recommendations whose similarity to the anchor falls below each $\tau$.

To measure practical impact, we define the \textbf{rent-flow rate} as the ratio between the number of recommendations that resulted in a rent-flow (user actions indicating strong intent, such as booking an in-person visit or submitting a rental offer) and the total number of recommendations shown. Focusing on rent flows provides a more accurate reflection of genuine user interest, since lower-intent signals like clicks tend to be noisy.

\subsection{Results and Discussion}

\noindent The single-resolution two-tower model (H3 resolution~9) increases Information Abundance by {\bfseries 158\%} (from 6.0 to 15.5) and Rent-Flow uplift by {\bfseries 84\%} (from 3.2\% to 5.9\%), emphasizing the advantage of the two-tower architecture. When adding multi-resolution embeddings, the two-tower model further improves Information Abundance by {\bfseries 40\%} (from 15.5 to 21.7) and Rent-Flow uplift by {\bfseries 85\%} (from 5.9\% to 10.9\%), demonstrating that richer spatial contexts significantly enhance both embedding quality and downstream usability.

\begin{table}[ht]
  \centering
  \caption{Embedding quality and downstream business impact}
  \label{tab:embed_quality}
  \begin{tabular}{lcc}
    \toprule
    Model & IA & Avg Rent-Flow uplift (\%) \\
    \midrule
    Matrix Factorization
      & 6.0
      & 3.2 \\
    two-tower (H3 Resolution 9)
      & \underline{15.5}
      & \underline{5.9} \\
    two-tower Multi-Resolution
      & \textbf{21.7}
      & \textbf{10.9} \\
    \bottomrule
  \end{tabular}
  \\[1ex]
  \footnotesize
  Information Abundance (IA) and simulated Average Rent-Flow uplift when re‐ranking house search results.
\end{table}

\begin{figure}[h]
  \centering
  \includegraphics[width=\linewidth]{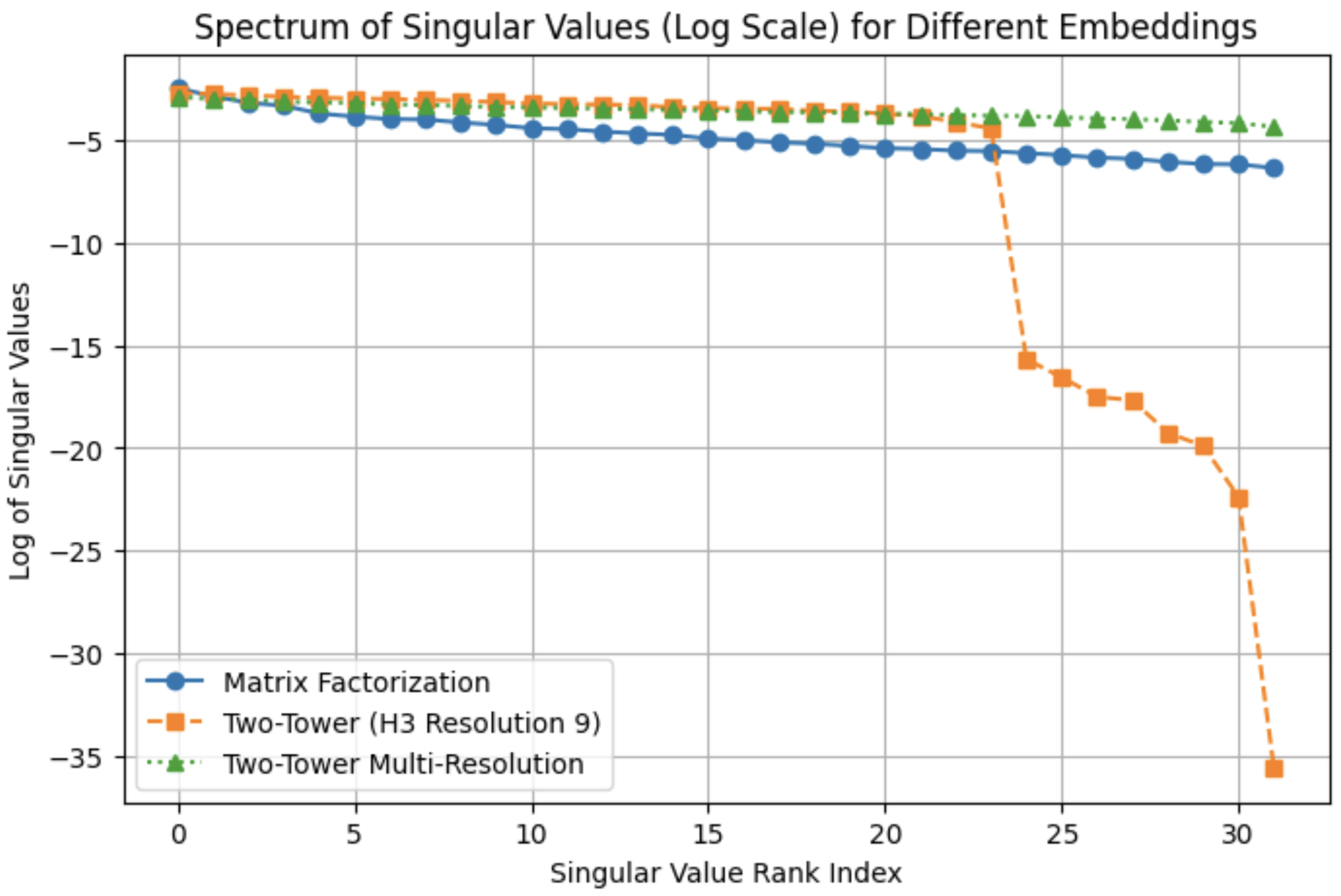}
  \caption{Logarithm of singular values ($\sigma_1$, sorted descending) for each model’s normalized embedding covariance matrix.}
  \Description{This image shows the logarithm of singular values plotted against their rank index. The x-axis is labeled "Singular Value Rank Index," and the y-axis is labeled "Log of Singular Values." There are three lines representing different models:

1. Matrix Factorization: Represented by blue circles connected by a solid line. This line starts at a higher value and remains relatively flat across the rank indices.

2. Two-tower (H3 Resolution 9): Represented by orange squares connected by a dashed line. This line starts similarly to the blue line but begins to drop sharply after a certain point, indicating a faster decay of singular values.

3. Two-tower Multi-Resolution: Represented by green triangles connected by a dotted line. This line also starts high and remains flat, similar to the blue line.
}
  \label{info-abundance}
\end{figure}

Figure~\ref{info-abundance} plots the singular-value spectral of each model’s normalized embedding covariance matrix. The multi-resolution model exhibits the flattest decay in $\log(\sigma_i)$, indicating that variance is distributed more evenly across dimensions.

\begin{figure}[h]
  \centering
  \includegraphics[width=\linewidth]{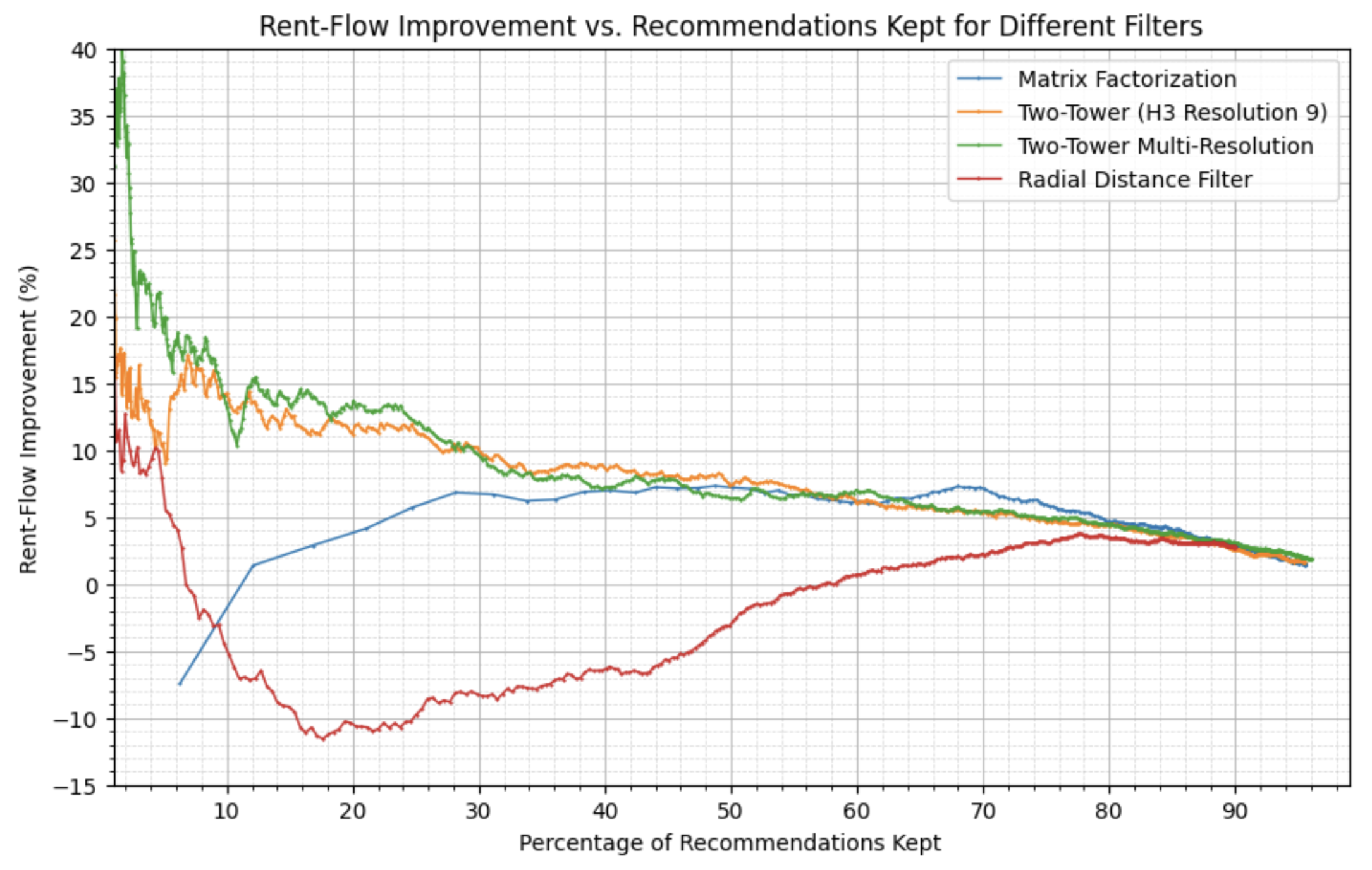}
  \caption{Simulated Rent-Flow Improvement. Percentage uplift in average rental-flow events (visits + offers) when re-ranking with each geo-aware model.}
  \Description{
  This plot shows the improvement in rent-flow when using different geo-aware models to re-rank recommendations. The x-axis represents the percentage of recommendations kept, while the y-axis shows the percentage improvement in rent-flow. Four lines represent different models: Matrix Factorization, two-tower (H3 Resolution 9), two-tower Multi-Resolution and a simple Radial Distance Filter.
}
  \label{rentflow}
\end{figure}

Figure~\ref{rentflow} shows the simulated rent-flow improvement across different recommendation filtering methods. Each curve represents how the relative improvement in rent-flow events (e.g., property visits and rental inquiries) changes as recommendations are progressively pruned based on embedding similarity thresholds. The horizontal axis indicates the percentage of original recommendations retained after filtering, while the vertical axis shows the corresponding percentage uplift in rent-flow compared to the baseline item-item recommender without geo-embedding-based filtering. Higher curves indicate better performance, demonstrating that embedding-based methods (particularly the two-tower Multi-Resolution model) effectively identify more relevant properties, resulting in a substantial increase in expected conversion.

We find that enriching the location encoder with both city-level and multi-scale H3 embeddings substantially boosts representation quality. This enhancement translates into promising increases in business-relevant metrics when applied to downstream recommendation tasks.

\section{Conclusion}

In this paper, we tackled the challenge of geographic sparsity in housing recommendations. We compared three approaches: a classical matrix factorization model, a two-tower architecture using single-resolution geographic features, and a two-tower model enhanced with multi-resolution H3 embeddings.

Our experiments show that incorporating multi-resolution geo-embeddings improves the system’s ability to represent locations and align recommendations with user intent.

As future work, we plan to validate these promising offline results through an online A/B test. We further seek to evaluate adding richer location attributes, for example, distance to points of interest such as metro or police stations. We will also explore refining the granularity and dynamics of geo-representations, and extending our approach to other real-estate recommendation scenarios.

\section{Authors}

Ivo is a Data Scientist in Recommendations at QuintoAndar, where he focuses on building and scaling personalized ranking systems. Before joining QuintoAndar, he spent seven years at a luxury fashion ecommerce company, developing end-to-end recommendation models, running large-scale A/B tests, and productionizing ML pipelines. \\

Guilherme Bonaldo is a Senior Data Scientist at QuintoAndar with eight years of experience in machine learning and AI, focusing on MLOps, credit risk prediction and search ranking models. Before joining QuintoAndar, he served as a consultant on machine learning projects across diverse sectors. He holds a B.S. in Chemical Engineering from the University of São Carlos (UFSCar).  \\

Pedro Nogueira is a Data Science Manager at QuintoAndar working in ranking and recommendations. Previously he also tackled Information Retrieval problems at a luxury fashion ecommerce company. With more than 10 years experience under his belt, Pedro’s passion is to build machine learning systems that have real impact in companies. He holds an MSc in Electrical and Computer Engineering from University of Porto.

\bibliographystyle{ACM-Reference-Format}
\bibliography{sample-base}

\end{document}